\documentclass[aps,prd,groupedaddress,showpacs,eqsecnum]{revtex4}
\usepackage{graphicx}% Include figure files
\usepackage{dcolumn}% Align table columns on decimal point
\usepackage{bm}% bold math
\begin{document}

\title{Role of $a_1(1260)$ resonance in multipion decays of light vector mesons.}

\author{N.~N.~Achasov}
\email[]{achasov@math.nsc.ru}
\author{A.~A.~Kozhevnikov}
\email[]{kozhev@math.nsc.ru}
%\thanks{}
\altaffiliation{}
\affiliation{Laboratory of  Theoretical Physics, S.~L.~Sobolev Institute for Mathematics,
630090, Novosibirsk, Russian Federation}

\date{\today}
\begin{abstract}
The contribution of the $a_1(1260)$ meson to the amplitudes of the
decays $\rho(770)\to4\pi$,  $\omega(782)\to5\pi$, and
$\phi(1020)\to5\pi$ is  analyzed  in the chiral model of
pseudoscalar, vector, and axial vector mesons based on the
generalized hidden local symmetry  added with the anomalous terms.
The analysis shows that inclusion of $a_1$ meson in the
intermediate states  results in enhancement of the branching
ratios of the above decays by the factor ranging from 1.3 to 1.9
depending on the mass of $a_1$ meson ranging from 1.23 GeV to
$m_{a_1}=m_\rho\sqrt{2}=1.09$ GeV, the greater factor standing in
case of lower mass of the $a_1$.
\end{abstract}
\pacs{11.30.Rd;12.39.Fe;13.30.Eg}

\maketitle

\section{Introduction}
\label{sec1}

In the low-energy domain, quantum chromodynamics (QCD) manifests
as an effective theory formulated in terms of colorless degrees of
freedom \cite{weinberg79}. They are introduced on the basis of
chiral $G=U(3)_L\times U(3)_R$ symmetry of the QCD lagrangian with
approximately massless $u$, $d$, and $s$ quarks. This symmetry is
supposed to be  spontaneously broken to $H=SU(3)_{R+L}$. As is
well known, the spontaneous symmetry breaking \cite{nambu} is
followed by the appearance of massless bosons \cite{goldstone}, in
the present case, nine pseudoscalar mesons $\pi^\pm$, $\pi^0$,
$K^\pm$, $K^0$, $\bar K^0$, $\eta$, and $\eta^\prime$. Their
effective lagrangian, including the interaction terms, is  fixed
by the symmetry breaking pattern $G\to H$, according to which the
fields of Goldstone bosons are treated as the coordinates in the
space $G/H$ \cite{weinberg68,ccwz}. Adding the Wess-Zumino term
\cite{wz} to the effective lagrangian removes the spurious
selection rule which forbids the processes with odd number of
Goldstone bosons.

There are several models which incorporate  the low-lying  vector
mesons $\rho(770)$, $\omega(782)$, $\phi(1020)$ etc.  into the
chiral theory, see Refs.~
\cite{weinberg68,schechter84,meissner88,bando88,birse96}. As far
as non-anomalous sector is concerned, there is the equivalence of
such models, see \cite{meissner88,birse96,harada03}. However, the
anomalous couplings are most conveniently incorporated into chiral
theory in the framework of approach based on the hidden local
symmetry (HLS) \cite{bando88,bando85}. The above vector mesons are
the gauge bosons of HLS. In particular, the convenience of HLS
rests on  the fact that  $\rho$, $\omega$, and $\phi$ mesons can
be accounted for without violation of the low energy theorems
\cite{bando88,harada03}. To avoid such a violation, other chiral
models of vector and pseudoscalar mesons rely essentially on the
subtraction to the gauged Wess-Zumino term \cite{wz}. The question
of the validity of each specific model is acute because in the
well-studied decays $\rho^0\to\pi^+\pi^-$,
$\omega\to\pi^+\pi^-\pi^0$ the final pions are not soft enough to
use the decay amplitudes in the tree approximation. On the other
hand, the multipion decays of vector mesons
$\rho^0(770)\to\pi^+\pi^-\pi^+\pi^-$, $\pi^+\pi^-\pi^0\pi^0$
\cite{rittenberg69,bramon93,kuraev95,plant96,ach00},
$\rho^\pm(770)\to\pi^\pm\pi^\pm\pi^\mp\pi^0$,
$\pi^\pm\pi^0\pi^0\pi^0$ \cite{ach00} and
$\omega(782),\phi(1020)\to\pi^+\pi^+\pi^-\pi^-\pi^0$,
$\pi^+\pi^-\pi^0\pi^0\pi^0$ \cite{ach03}, where the final pions
are truly soft to rely on the lowest order Born amplitudes, can be
good candidates for testing the chiral models of vector and
pseudoscalar mesons \cite{ach00,ach03}. A brief accounts of the
$\omega\to5\pi$ and $\phi\to5\pi$ results are given, respectively,
in Refs.~\cite{ach00a,ach04}.

In Refs.~\cite{ach00,ach03} devoted to the evaluation of the
branching ratios of the above multipion decays, we neglected the
contribution of the axial vector $a_1(1260)$ meson. The present
paper  addresses the question to what extent the inclusion  of
this resonance affects the branching ratios of the decays listed
above. As is known, chiral models admit the contribution of the
axial vector mesons like $a_1(1260)$, see reviews
\cite{meissner88,bando88}.  We shall use the generalized hidden
local symmetry model (GHLS) \cite{bando88a} because it accounts
for the contributions of the vector and axial vector resonances in
a most elegant way.

The material of the paper is organized as follows. In
sec.~\ref{chirlagr}, starting from the GHLS lagrangian
\cite{bando88a}, the lagrangian of $\pi$, $\rho$, $\omega$, and
$a_1$ mesons is obtained at the lowest number of derivatives
necessary for the derivation of the $\rho\to4\pi$,
$\omega\to5\pi$, and $a_1\to3\pi$ decay amplitudes.
Sec.~\ref{a1width} is devoted to the derivation of the
$a_1\to3\pi$ decay amplitude, with the emphasis on its  behavior
at the vanishing pion momenta. Using the derived expression, the
$a_1\to3\pi$ decay width is evaluated assuming different  masses
of the $a_1$ meson. The contribution of the $a_1(1260)$ resonance
to the $\rho\to4\pi$ decay amplitude is found in
sec.~\ref{rho4pi}. Its influence on the $\omega\to5\pi$ and
$\phi\to5\pi$ decay amplitudes is discussed in the same section.
The results of the evaluation of the branching ratios of the
decays $\rho\to4\pi$, $\omega\to5\pi$, and $\phi\to5\pi$, taking
into account the contributions of the $a_1$ meson and the
additional $\rho\rho\pi\pi$ vertex Eq.~({\ref{rhorhopipi}),  are
presented in sec.~\ref{results}. Sec.~\ref{discussion} is devoted
to a brief discussion of the results obtained in the present
paper.

\section{Chiral invariant lagrangian of $\pi$, $\rho$, $\omega$, and $a_1$ mesons
with lowest number of  derivatives} \label{chirlagr} ~

The basis of the derivation is the lagrangian of the generalized
hidden local symmetry model \cite{bando88a} (GHLS) which, in the
gauge $\xi_M=1$, $\xi^\dagger_L=\xi_R=\xi$, looks as
\begin{eqnarray}
{\cal L}^{({\rm GHLS})}&=&a_0f^2_\pi{\rm
Tr}\left(\frac{\partial_\mu\xi^\dagger\xi+\partial_\mu\xi\xi^\dagger}{2i}-gV_\mu\right)^2+
b_0f^2_\pi{\rm
Tr}\left(\frac{\partial_\mu\xi^\dagger\xi-\partial_\mu\xi\xi^\dagger}{2i}+gA_\mu\right)^2
+\nonumber\\ &&c_0f^2_\pi g^2{\rm Tr}A^2_\mu+d_0f^2_\pi{\rm
Tr}\left(\frac{\partial_\mu\xi^\dagger\xi-\partial_\mu\xi\xi^\dagger}{2i}\right)^2
-\frac{1}{2}{\rm
Tr}\left(F^{(V)2}_{\mu\nu}+F^{(A)2}_{\mu\nu}\right)-\nonumber\\&&
i\alpha_4g{\rm Tr}[A_\mu,A_\nu]F^{(V)}_{\mu\nu}+2i\alpha_5g{\rm
Tr}\left(\left[\frac{\partial_\mu\xi^\dagger\xi-\partial\xi\xi^\dagger}{2ig},A_\nu\right]
+[A_\mu,A_\nu]\right)F^{(V)}_{\mu\nu}. \label{lghls}\end{eqnarray}
The notations, assuming the restriction to the sector of the
non-strange mesons,  are
\begin{eqnarray}
F^{(V)}_{\mu\nu}&=&\partial_\mu V_\nu-\partial_\nu
V_\mu-i[V_\mu,V_\nu]-i[A_\mu,A_\nu],\nonumber\\
F^{(A)}_{\mu\nu}&=&\partial_\mu A_\nu-\partial_\nu
A_\mu-i[V_\mu,A_\nu]-i[A_\mu,V_\nu],\nonumber\\
V_\mu&=&\left(\frac{{\bm\tau}}{2}\cdot{\bm\rho}_\mu\right)+\frac{1}{2}\omega_\mu,\nonumber\\
A_\mu&=&\left(\frac{{\bm\tau}}{2}\cdot{\bm A
}_\mu\right),\nonumber\\ \xi&=&\exp
i\frac{{\bm\tau}\cdot{\bm\pi}}{2f_\pi}, \label{not}\end{eqnarray}
where ${\bm\rho}_\mu$, $\omega_\mu$, ${\bm\pi}$ are the vector
meson $\rho$, $\omega$  and pseudoscalar pion fields,
respectively, ${\bm A}_\mu$ is the axial vector field (not $a_1$
meson!), ${\bm\tau}$ is the isospin Pauli matrices, $f_\pi$=92.4
MeV is the pion decay constant, $[,]$ stands for commutator.
Hereafter the boldface characters, cross ($\times$), and dot
($\cdot$) stand for vectors, vector product, and scalar product,
respectively, in the isotopic space. The constants $a_0$, $b_0$,
$c_0$, $d_0$, $\alpha_{4,5}$ are specified below. The relevant
terms of the lagrangian describing $a_1$ meson and its couplings
to the $\rho\pi$ and $3\pi$ systems can be obtained from
Eq.~(\ref{lghls}) following the steps \cite{bando88a} outlined
below. First,  we exclude the  mixing term
\begin{equation}
\Delta{\cal L}^{(a_1-\pi)}\propto{\rm
tr}A_\mu\frac{\partial_\mu\xi^\dagger\xi-\partial_\mu\xi\xi^\dagger}{2i}
\propto{\bm
A}_\mu\left(\partial_\mu{\bm\pi}+\frac{1}{6f^2_\pi}[{\bm\pi}\times[{\bm\pi}
\times\partial_\mu{\bm\pi}]]+\cdots\right)
\label{genmix}\end{equation} by introducing the field of $a_1$
meson $$a_\mu=\left(\frac{{\bm\tau}}{2}\cdot{\bm a}_\mu\right)$$
as follows:
\begin{equation}
A_\mu=a_\mu-\frac{b_0}{g(b_0+c_0)}\frac{\partial_\mu\xi^\dagger\xi-\partial_\mu\xi\xi^\dagger}
{2i}. \label{a1field}\end{equation} Note that in distinction with
Ref.~\cite{bando88a}, where the  mixing term  ${\bm
A}_\mu\partial_\mu{\bm\pi}$ with the lowest order derivative in
pion field  is rotated away, we do so with the entire nonlinear
combination Eq.~(\ref{genmix}). We postpone the justification of
our choice until discussing the $3\pi$ decay width of $a_1$ meson
in sec.~\ref{a1width}.  The above diagonalization introduces the
unwanted momentum dependence of the $\rho\pi\pi$ vertex, which can
be cancelled by the counter terms \cite{bando88a}. They are
represented  by the terms  containing the parameters
$\alpha_{4,5}$ in  Eq.(\ref{lghls}). Following
Ref.~\cite{bando88a}, we retain only the terms with
$-\alpha_4=\alpha_5=\alpha_6\not=0$, with the further fixing
$-\alpha_4=\alpha_5=1$ \cite{bando88a}. The second step is the
renormalization $f_\pi\to Z^{-1/2}f_\pi$, ${\bm\pi}\to
Z^{-1/2}{\bm\pi}$, $(a_0,b_0,c_0,d_0)=Z(a,b,c,d),$ where
$$\left(d_0+\frac{b_0c_0}{b_0+c_0}\right)Z^{-1}=1.$$ The last step
is the choice \cite{bando88a} $a=b=c=2$, $d=0$ which results in
the universality $g_{\rho\pi\pi}=g$ and vector dominance of the
$\rho\pi\pi$ coupling, the Kawarabayashi-Suzuki-Riazzuddin-
Fayyazuddin (KSRF) relation \cite{ksrf}
\begin{equation}
{2g^2_{\rho\pi\pi}f^2_\pi\over m^2_\rho}=1, \label{ksrf}
\end{equation}
and the Weinberg relation \cite{weinberg67}
\begin{equation}
m_{a_1}=\sqrt{2}m_\rho=1.09\mbox{
GeV},\label{massa1}\end{equation}see Eq.~(\ref{lnan}) and
(\ref{a1rhopi}). The $\rho\pi\pi$ coupling constant resulting from
Eq.~(\ref{ksrf}) is $g_{\rho\pi\pi}=5.9$. Finally, using the weak
field expansion
\begin{eqnarray}
\frac{\partial_\mu\xi^\dagger\xi-\partial_\mu\xi\xi^\dagger}{2i}&=&
-\frac{1}{2f_\pi}{\bm\tau}\cdot\left(\partial_\mu{\bm\pi}+\frac{1}{6f^2_\pi}[{\bm\pi}\times[{\bm\pi}
\times\partial_\mu{\bm\pi}]]+\cdots\right),\nonumber\\
\frac{\partial_\mu\xi^\dagger\xi+\partial_\mu\xi\xi^\dagger}{2i}&=&
-\frac{1}{4f^2_\pi}{\bm\tau}\cdot[{\bm\pi}\times\partial_\mu{\bm\pi}]
\left(1-\frac{{\bm\pi}^2}{12f^2_\pi}\cdots\right),
\label{weakexp}\end{eqnarray} one obtains  the following
lagrangian of the $\rho$, $\omega$, $a_1$, and $\pi$ mesons  at
the order required for the evaluation of the $a_1$ meson
contribution to the decay $\rho\to4\pi$: $${\cal L}^{({\rm
GHLS})}\approx{\cal L}^{({\rm HLS})}+\Delta{\cal L}^{({\rm
GHLS})},$$ where
\begin{eqnarray}
\cal{L}^{\rm
HLS}&=&-{1\over4}\bm{\rho}_{\mu\nu}^2-\frac{1}{4}\omega_{\mu\nu}^2+
{1\over2}ag^2f_\pi^2\left(\bm{\rho}^2_\mu+\omega_\mu^2\right)+
{1\over2}\left(\partial_\mu\bm{\pi}\right)^2-{1\over2}m^2_\pi\bm{\pi}^2+
{m^2_\pi\over24f^2_\pi}\bm{\pi}^4+   \nonumber\\
&&{1\over2f^2_\pi}\left({a\over4}-{1\over3}\right)
[\bm{\pi}\times\partial_\mu \bm{\pi}]^2+
{1\over2}ag\left(1-{\bm{\pi}^2\over12f^2_\pi}\right)\left(\bm{\rho}_\mu\cdot
[\bm{\pi}\times\partial_\mu\bm{\pi}]\right) \label{lnan}
\end{eqnarray}
is the weak field limit of the lagrangian of HLS including the
terms $\propto m^2_\pi$ which explicitly  violate the chiral
symmetry,
\begin{eqnarray}
\bm{\rho}_{\mu\nu}&=&\partial_\mu\bm{\rho}_\nu-\partial_\nu\bm{\rho}_\mu+
g[\bm{\rho}_\mu\times\bm{\rho}_\nu],\nonumber\\
\omega_{\mu\nu}&=&\partial_\mu\omega_\nu-\partial_\nu\omega_\mu
\label{strength}
\end{eqnarray}
are  the field strengths of the isovector  $\bm{\rho}_\mu$ and
isoscalar $\omega_\mu$ fields, $g$ is the gauge coupling constant,
$a=2$ is HLS parameter. As is clear from Eq.~(\ref{lnan}),
\begin{eqnarray}
g_{\rho\pi\pi}&=&{1\over2}ag,   \nonumber\\
m_\rho^2&=&ag^2f_\pi^2\label{rhoparam}\end{eqnarray} are the
$\rho\pi\pi$ coupling constant and the $\rho$ mass squared,
respectively. Note that $m_\omega=m_\rho$ in HLS.  The lagrangian
$$\Delta{\cal L}^{({\rm GHLS})}={\cal L}^{(a_1\rho\pi)}+{\cal
L}^{(\rho\rho\pi\pi)}+{\cal L}^{(4\pi)}$$ is  the contribution of
that part of the GHLS lagrangian Eq.~(\ref{lghls}) which contains
the axial vector field $A_\mu$, the terms originating from the
diagonalization of $A-\pi$ mixing Eq.~(\ref{genmix}), and the
counter terms. It consists of the terms responsible for the free
$a_1$ field and its interaction with the $\rho\pi$ and $3\pi$
states,
\begin{eqnarray}
{\cal L}^{(a_1\rho\pi)}&=&-\frac{1}{4}(\partial_\mu{\bm a}_\nu
-\partial_\nu{\bm a}_\mu)^2+\frac{1}{2}(b+c)g^2f^2_\pi{\bm
a}_\mu^2-\frac{1}{f_\pi}(\partial_\mu{\bm\rho}_\nu
-\partial_\nu{\bm\rho}_\mu)\cdot[{\bm
a}_\mu\times\partial_\nu{\bm\pi}]-\nonumber\\&&\frac{1}{2f_\pi}(\partial_\mu{\bm
a}_\nu -\partial_\nu{\bm a}_\mu)\cdot[{\bm
\rho}_\mu\times\partial_\nu{\bm\pi}]-\frac{1}{8gf^3_\pi}[{\bm
a}_\mu\times\partial_\nu{\bm\pi}]\cdot[\partial_\mu{\bm\pi}\times\partial_\nu{\bm\pi}]
-\nonumber\\&&\frac{1}{4gf^3_\pi}\partial_\mu{\bm
a}_\nu\cdot[{\bm\pi}\times[\partial_\mu{\bm\pi}\times\partial_\nu{\bm\pi}]],
\label{a1rhopi}\end{eqnarray}the term describing the
$\rho\rho\pi\pi$ and the higher derivative point-like
$\rho\to4\pi$ vertex vertices
\begin{equation}
{\cal L}^{(\rho\rho\pi\pi)}=-\frac{1}{16f^2_\pi}\left([{\bm
\rho}_\mu\times\partial_\nu{\bm\pi}]-[{\bm
\rho}_\nu\times\partial_\mu{\bm\pi}]\right)^2-\frac{1}{8gf^4_\pi}[{\bm
\rho}_\mu\times\partial_\nu{\bm\pi}]\cdot[{\bm\pi}\times[\partial_\mu{\bm\pi}
\times\partial_\nu{\bm\pi}]], \label{rhorhopipi}\end{equation} and
the higher derivative $4\pi$ vertex: \begin{equation}{\cal
L}^{(4\pi)}=\frac{1}{64g^2f^4_\pi}[\partial_\mu{\bm\pi}\times\partial_\nu{\bm\pi}]^2.
\label{high4pi}\end{equation} Note that when deriving the above
lagrangians,  we have not used the equation of motion of the
fields $\pi$, $\rho$, and $a_1$. One should have in mind that in
the decays of our interest the final pions are non-relativistic,
$p_\mu\approx(m_\pi,0,0,0)$. The direct calculation shows that the
ratio of the contribution from Eq.~(\ref{high4pi}) to the lowest
derivative $\pi\pi$ scattering amplitude  is about
$(m_\pi/4gf_\pi)^2\approx4\times10^{-3}$, in agreement with the
expectations of the chiral perturbation theory. Hence, we shall
ignore this contribution in what follows. In the meantime, the
higher derivative point-like vertex $\rho\to4\pi$ in
Eq.~(\ref{rhorhopipi}) cannot be omitted, because it is essential
for validity of the Adler condition for the contribution to the
$\rho\to4\pi$ decay amplitude originating from
Eq.~(\ref{rhorhopipi}). See details in sec.~\ref{rho4pi}.

The terms of the effective lagrangian necessary for the
calculation of  the $\omega\to5\pi$ decay amplitude are obtained
from the weak field limit of the terms \cite{bando88,harada03}
induced by the anomalous term of Wess and Zumino \cite{wz}. The
corresponding expression looks as
\begin{eqnarray}
\cal{L}^{\rm
an}&=&{n_cg\over32\pi^2f^3_\pi}(c_1-c_2-c_3)\varepsilon_{\mu\nu\lambda\sigma}\omega_\mu
\left(\partial_\nu\bm{\pi}\cdot[\partial_\lambda\bm{\pi}\times\partial_\sigma\bm{\pi}]
\right)+        \nonumber\\ &&
{n_cg\over128\pi^2f^5_\pi}\left[-c_1+{5\over3}\left(c_2+c_3\right)\right]
\varepsilon_{\mu\nu\lambda\sigma}\omega_\mu
\left(\partial_\nu\bm{\pi}\cdot[\partial_\lambda\bm{\pi}\times\partial_\sigma\bm{\pi}]
\right)\bm{\pi}^2-        \nonumber\\ &&
{n_cg^2c_3\over8\pi^2f_\pi}\varepsilon_{\mu\nu\lambda\sigma}\partial_\mu\omega_\nu
\left\{\left(\bm{\rho}_\lambda\cdot\partial_\sigma\bm{\pi}\right)+{1\over6f^2_\pi}
\left[\left(\bm{\rho}_\lambda\cdot\bm{\pi}\right)\left(\bm{\pi}\cdot\partial_\sigma
\bm{\pi}\right)-\bm{\pi}^2\left(\bm{\rho}_\lambda\cdot\partial_\sigma\bm{\pi}\right)
\right]\right\}-     \nonumber\\ &&
{n_cg^2\over8\pi^2f_\pi}(c_1+c_2-c_3)\varepsilon_{\mu\nu\lambda\sigma}\omega_\mu
\left\{\frac{1}{4f^2_\pi}\left(\partial_\nu\bm{\pi}\cdot\bm{\rho}_\lambda\right)\left(\bm{\pi}\cdot
\partial_\sigma\bm{\pi}\right)-\right.\nonumber\\&&\left.
{g\over4}\left(\left[\bm{\rho}_\nu\times
\bm{\rho}_\lambda\right]\cdot\partial_\sigma\bm{\pi}\right)\right\}
\label{lan}
\end{eqnarray}
\cite{ach03}, where $n_c=3$ is the number of colors, $c_{1,2,3}$
are arbitrary constants multiplying three independent structures
in the solution \cite{bando88,harada03} of the Wess-Zumino anomaly
equation \cite{wz}. The  normalization of $c_{1,2,3}$ is in accord
with Ref.~\cite{harada03}. As is evident from Eq.~(\ref{lan}), the
$\omega\rho\pi$ coupling constant is
\begin{equation}
g_{\omega\rho\pi}=-{n_cg^2c_3\over8\pi^2f_\pi}.
\label{gomrp}\end{equation} Assuming in what follows the relation
\begin{equation}
c_1-c_2-c_3=0,\label{rel1}\end{equation} i.e. the absence of the
point like $\omega\to\pi^+\pi^-\pi^0$ amplitude \cite{fn1}, and
using the $\omega\to\pi^+\pi^-\pi^0$ partial width to extract
$g_{\omega\rho\pi}$, the $\rho\to\pi^+\pi^-$ partial width and
Eq.~(\ref{rhoparam}) to extract $g=g_{\rho\pi\pi}\approx6$
(assuming $a=2$), one finds $c_3\approx1$. Hereafter we use the
particle parameters (masses, full and partial widths etc.) taken
from Ref.~\cite{pdg}. The decay $\phi\to5\pi$ is described by the
effective lagrangian similar to Eq.~(\ref{lan}), see
Ref.~\cite{ach03}. The  evaluation of the branching ratios of the
decays $\omega,\phi\to5\pi$ with the neglect of the $a_1$ meson
and counter term contributions  is performed in Ref.~\cite{ach03}.

\section{The width of $a_1$ resonance in GHLS }
\label{a1width} ~

Let us find the width of the decay $a_1\to3\pi$ in GHLS. This task
is necessary, because the original Ref.~\cite{bando88a} contains
only the discussion of the $a_1\to\rho\pi$ decay width which, as
it will be clear, overestimates the true $a_1\to3\pi$ decay width.
When so doing, the point-like $a_1\to3\pi$ vertex is essential.
The amplitude of, say, the decay $a^0_1\to\pi^+\pi^-\pi^0$ can be
found from the lagrangian Eq.~(\ref{a1rhopi}):
\begin{equation}
iM(a^0_{1Q}\to\pi^+_{q_1}\pi^-_{q_2}\pi^0_{q_3})=\epsilon_\mu
J_\mu(a^0_{1Q}\to\pi^+_{q_1}\pi^-_{q_2}\pi^0_{q_3}),\label{ampa13pi}\end{equation}
where $\epsilon$ stands for the four-vector of polarization of the
$a_1$ meson,  particles  are labelled by their four-momenta, and
\begin{eqnarray}
J_\mu(a^0_{1Q}\to\pi^+_{q_1}\pi^-_{q_2}\pi^0_{q_3})&=&\frac{ag}{4f_\pi}(1+
P_{12})(1-P_{13})\left\{\frac{1}{D_\rho(q_1+q_3)} \left[2(1-
P_{13})q_{1\mu}(q_2q_3)+\right.\right.\nonumber\\&&\left.\left.(1-
P_{12})q_{1\mu}(Qq_2)\right]-\frac{1}{2m^2_\rho}
q_{1\mu}[2(Qq_3)+(q_2q_3)]\right\}. \label{ja13pi}\end{eqnarray}
Hereafter $P_{ij}$ is the operator that interchanges the
four-momenta $q_i$ and $q_j$,
\begin{equation}
D_\rho(k)=m^2_\rho-k^2-i\sqrt{k^2}\Gamma_\rho(\sqrt{k^2})\label{proprho}\end{equation}
is the inverse  propagator of the $\rho$ meson whose energy
dependent width above the $\pi^+\pi^-$ threshold and below the
$K^\ast\bar K$ one includes the $\pi^+\pi^-$, $K\bar K$, and
$\omega\pi$ decay modes:
\begin{equation}
\Gamma_\rho(\sqrt{k^2})=\frac{g^2_{\rho\pi\pi}}{6\pi
k^2}q^3_{\pi\pi}(k^2)+\frac{g^2_{\rho K\bar K}}{3\pi
k^2}q^3_{K\bar
K}(k^2)\theta(\sqrt{k^2}-2m_K)+\frac{g^2_{\rho\omega\pi}}{12\pi}q^3_{\omega\pi}(k^2)
\theta(\sqrt{k^2}-m_\omega-m_\pi).\label{rhowidth}\end{equation}
Here $\theta$ is the usual step function, while
\begin{equation}
q_{ab}(k^2)=\frac{1}{2\sqrt{k^2}}\sqrt{[k^2-(m_a+m_b)^2][k^2-(m_a-m_b)^2]}
\label{momemtum}\end{equation} is the momentum of the final state
particle in the rest frame system of the decaying particle. In
case of energies $E\sim m_\phi$ discussed in the present paper,
only the $\pi^+\pi^-$ decay mode is essential. In the quark model,
the coupling constants are related in the following way:
$g^2_{\rho K\bar K}=\frac{1}{2}g^2_{\phi K\bar K}$,
$g_{\rho\omega\pi}=g_{\omega\rho\pi}$, where $g_{\phi K\bar K}$ is
calculated from the $\phi\to K\bar K$ decay width.

One can convince oneself that the expression (\ref{ampa13pi})
vanishes at the vanishing four-momentum of any final pion. This
property called  the Adler condition,  expresses the chiral
invariance of the underlying theory. [To be more precise, the
check based on the Adler condition hereafter is applied  in the
narrow $\rho$ width approximation. Indeed, it should be recalled
that the finite width effects are attributed to the loop
corrections which are beyond the tree approximation adopted in the
present paper. Numerically, at energies of our concern the
invariant mass of a pion pair is $m<0.6$ GeV, so that
$m\Gamma_\rho(m)/(m^2_\rho-m^2)<0.26$, and the effects of the
$\rho$ width  in the diagrams with the non resonant $\rho$ meson
are  small.] The $\rho$ pole contribution without the point-like
$a_1\to\pi^+\pi^-\pi^0$ vertex does not possess this property.
Remarkably, the Adler condition for the $a_1\to3\pi$ decay
amplitude Eq.(\ref{ampa13pi}) is valid even in the case of the
off-mass-shell $a_1$ meson. This is very useful because one can
safely add the $a_1$ contribution to the amplitudes which satisfy
the Adler condition, without spoiling this property.  At this
point, one can justify the choice of the diagonalization of the
$A-\pi$ lagrangian used in sec.~\ref{chirlagr}. Indeed, when the
$A-\pi$ mixing is excluded in the first order in the $\pi$ field,
it is equivalent to the adding the term
\begin{eqnarray}
i\Delta
M(a^0_{1Q}\to\pi^+_{q_1}\pi^-_{q_2}\pi^0_{q_3})&=&\frac{g}{3f_\pi}(\epsilon,q_1+q_2-2q_3)
+\frac{1}{4gf^3_\pi}\left[(\epsilon
q_3)(Q,q_1+q_2)-\right.\nonumber\\&&\left.(\epsilon,q_1+q_2)(Qq_3)\right]\label{a13piadd}
\end{eqnarray}
to the right hand side of Eq.~(\ref{ampa13pi}). Hereafter  $(a,b)$
stands for the Lorentz scalar product in cases when the
four-vectors $a$ or $b$ are sums of other four-vectors. As is
evident from Eq.~(\ref{a13piadd}), $i\Delta M$ does not vanish at
$q_1=0$ but rather reduces  to the  expression
$$iM(a^0_{1Q}\to\pi^+_{q_1=0}\pi^-_{q_2}\pi^0_{q_3})=\frac{g}{f_\pi}(\epsilon
q_2)\left(1-\frac{Q^2}{m^2_{a_1}}\right),$$ where
$m^2_{a_1}=4g^2f^2_\pi$, which vanishes only on the mass shell of
the $a_1$ meson. This would result in the breaking of the Adler
condition for the $\omega,\phi\to5\pi$ decay amplitudes upon
taking the $a_1$ resonance into account. In turn, it would demand
adding further counter terms, besides those proposed in
Ref.~\cite{bando88a}, to make the amplitude chirally invariant.
The above amplitudes with the neglect of the $a_1$ meson were
shown to obey the Adler condition \cite{ach03}.

The energy dependence of  the $a^0_1\to\pi^+\pi^-\pi^0$ decay
width can be found  from the expression
\begin{equation}
\Gamma_{a^0_1\to\pi^+\pi^-\pi^0}(m)=\frac{1}{3\times2^8\times\pi^3\times
m^3}\int^{(m-m_\pi)^2}_{4m^2_\pi} ds_1\int_{u_{1-}}^{u_{1+}}du_1
|M(a^0\to\pi^+_{q_1}\pi^-_{q_2}\pi^0_{q_3})|^2,\label{a13piwid}\end{equation}
where $|M(a^0\to\pi^+_{q_1}\pi^-_{q_2}\pi^0_{q_3})|^2$ is the
modulus squared of the amplitude Eq.~(\ref{ampa13pi}) summed over
three polarization states of the $a_1$. It should be expressed
through the invariant Kumar variables \cite{kumar}
$m^2=(q_1+q_2+q_3)^2$, $s_1=(q_2+q_3)^2$, $u_1=(q_1+q_3)^2$. The
limits of integration over $u_1$ are $$
u_{1\pm}=\frac{1}{2}(m^2+3m^2_\pi-s_1)\pm\frac{1}{2s_1}\sqrt{\lambda(s_1,m^2_\pi,m^2_\pi)
\lambda(m^2,s_1,m^2_\pi)},$$where
\begin{equation}
\lambda(x,y,z)=x^2+y^2+z^2-2xy-2xz-2yz.\label{lambda}\end{equation}
The results of the evaluation are shown in Fig.~\ref{width_a1}.
%%%%%%%%%%%%%%%%%%%%%%%%%%%%%%%%%%%%%%%%%%%%%%%%%%%%
\begin{figure}
\includegraphics{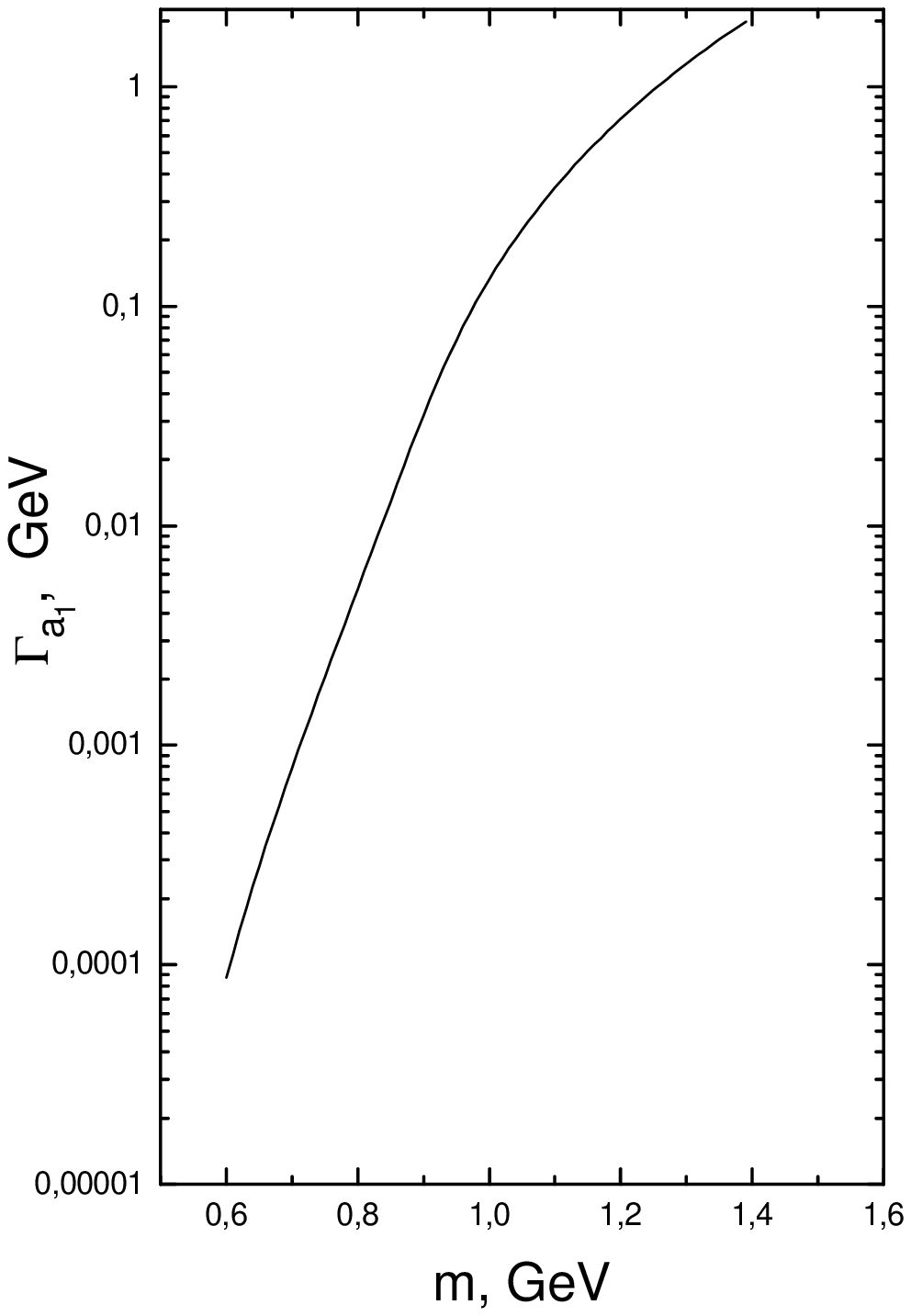}
\caption{\label{width_a1}The energy dependence of the
$a^0_1\to\pi^+\pi^-\pi^0$ decay width  calculated in the
generalized hidden local symmetry model.}
\end{figure}
%%%%%%%%%%%%%%%%%%%%%%%%%%%%%%%%%%%%%%%%%%%%%%%%%
The above width rises rapidly with  increasing $m$. In particular,
one obtains
$\Gamma_{a_1}\equiv\Gamma_{a^0_1\to\pi^+\pi^-\pi^0}(m)=320$,
$860$, $1024$ MeV at, respectively, $m=1090$, $1230$, $1260$ MeV.
For comparison, the $a_1\to\rho\pi$ decay width in the narrow
$\rho$ width approximation is 420, 1100, 1240 MeV, respectively.
Since  $\sqrt{s}\leq m_\phi=1020$ MeV is of our main concern, the
upper kinematical bound  of the invariant mass  of the three pion
system is 740 MeV. In the mass range $m\leq740$ MeV  the $a_1$
width is rather small, $\Gamma_{a_1}<1.7$ MeV, and can be safely
neglected.

\section{The $a_1$ and counter term contributions to the $\rho\to4\pi$  decay amplitude}
\label{rho4pi} ~

The $\rho\to4\pi$ decay amplitudes obtained in
Refs.~\cite{ach00,ach03} from the HLS lagrangian Eq.~(\ref{lnan}),
upon neglecting the $a_1$ meson contribution, obey the Adler
condition. In the GHLS approach, the additional terms originate,
first, from the lagrangian Eq.~(\ref{a1rhopi}) and are represented
by the diagram Fig.~\ref{diagr}(a), where, for each specific decay
$\rho^0\to\pi^+\pi^-\pi^+\pi^-$, $\pi^+\pi^-\pi^0\pi^0$,
$\rho^\pm\to\pi^\pm\pi^\pm\pi^\mp\pi^0$, $\pi^\pm\pi^0\pi^0\pi^0$,
one should take the sum of the diagrams with all possible
permutations of the final pion momenta. Second, there are the
terms which do not contain $a_1$ meson explicitly but result from
the exclusion  of the axial vector-pseudoscalar mixing term
Eq.~(\ref{genmix}). They are represented by the diagrams
Fig.~\ref{diagr}(b) and (c) and correspond to the first and second
term in the right hand side of Eq.~(\ref{rhorhopipi}). Again, one
should include the sum of the diagrams with  all possible
permutations of the final pion momenta.
%%%%%%%%%%%%%%%%%%%%%%%%%%%%%%%%%%%%%%%%%%%%%%%%%%%
\begin{figure}
\includegraphics{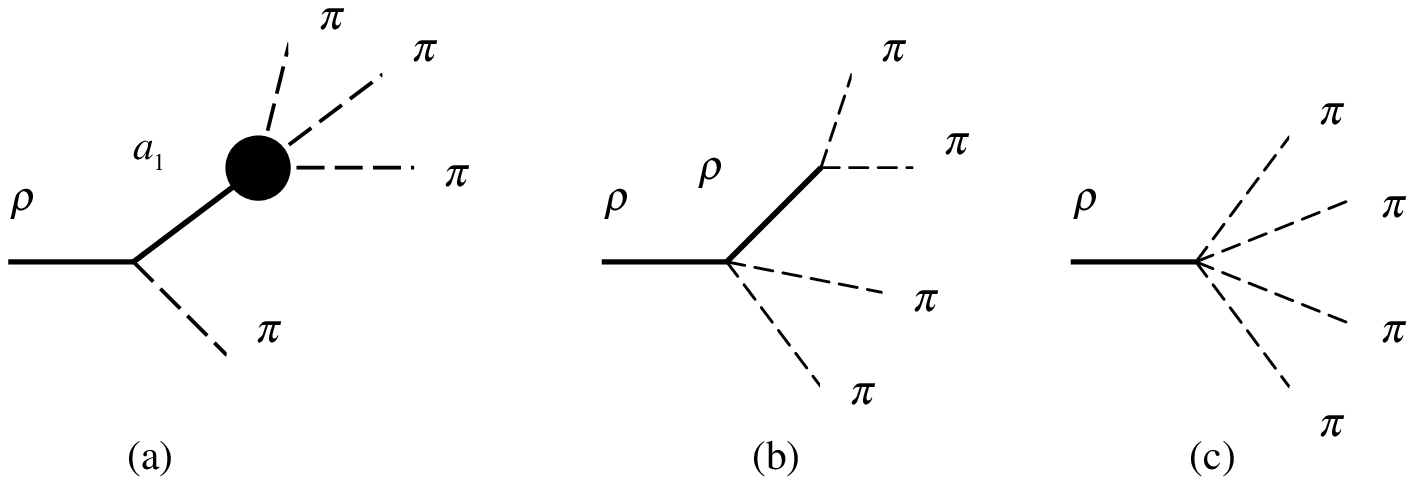}
\caption{\label{diagr}Diagrams corresponding to the contribution
of the   intermediate $a_1$ meson (a), $\rho\rho\pi\pi$ (b) and
point-like $\rho\to4\pi$ (c) vertices, respectively. The shaded
circle in (a) denotes the $a_1\to3\pi$ decay amplitude similar to
Eq.(\ref{ampa13pi}).}
\end{figure}
%%%%%%%%%%%%%%%%%%%%%%%%%%%%%%%%%%%%%%%%%%%%%%%%%
The $a_1$ contribution to the $\rho\to4\pi$ decay amplitude can be
obtained in the following way. When so doing, we present the
details for the $\rho^0\to\pi^+\pi^-\pi^+\pi^-$ decay mode only,
since other modes can be treated similarly. One has
\begin{eqnarray} i\Delta
M^{(a_1\rho\pi)}(\rho^0_q\to\pi^+_{q_1}\pi^+_{q_2}\pi^-_{q_3}\pi^-_{q_4})&=&(1+P_{34})J_\mu(\rho^0_q\to
a^+_1\pi^-_{q_4})\frac{i\left[\eta_{\mu\nu}-\frac{(q-q_4)_\mu(q-q_4)_\nu}{m^2_{a_1}}\right]}{D_{a_1}(q-q_4)}
\times\nonumber\\&&J_\nu(a^+_1\to\pi^+_{q_1}\pi^+_{q_2}\pi^-_{q_3})+(1+P_{12})J_\mu(\rho^0_q\to
a^+_1\pi^-_{q_2})\times\nonumber\\&&
\frac{i\left[\eta_{\mu\nu}-\frac{(q-q_2)_\mu(q-q_2)_\nu}{m^2_{a_1}}\right]}{D_{a_1}(q-q_2)}
J_\nu(a^-_1\to\pi^+_{q_1}\pi^-_{q_3}\pi^-_{q_4}),
\label{dela1}\end{eqnarray} where the inverse propagator of the
$a_1$ meson is
\begin{equation}
D_{a_1}(k)=m^2_{a_1}-k^2.\label{propa1}\end{equation}   The  decay
current is
\begin{equation}
J_\mu(\rho^0_q\to
a^+_1\pi^-_{q_4})=\frac{1}{2f_\pi}\left[(\epsilon
q_4)(2q-q_4)_\mu-(3qq_4-q^2_4)\epsilon_\mu\right],\label{jrhoa1pi}\end{equation}
$\epsilon$ is the polarization four-vector of the initial $\rho$
meson, and $J_\nu(a^+_1\to\pi^+_{q_1}\pi^+_{q_2}\pi^-_{q_3})$ is
obtained from  Eq.~(\ref{ja13pi}) by inverting the overall sign.
The expression for $J_\nu(\rho^0_q\to a^-_1\pi^+_{q_2})$ is
obtained from Eq.~(\ref{jrhoa1pi}) by inverting the sign while the
expression for $J_\nu(a^-_1\to\pi^+_{q_1}\pi^-_{q_3}\pi^-_{q_4})$
is obtained from
$J_\nu(a^+_1\to\pi^+_{q_1}\pi^+_{q_2}\pi^-_{q_3})$ upon the charge
conjugation followed by the replacements $q_1\longleftrightarrow
q_3$, $q_2\longleftrightarrow q_4$ of the final pion momenta. One
can directly show that the amplitude Eq.~(\ref{dela1}) obeys the
Adler condition at the vanishing of any pion momentum.

Next let us give the expressions for the contribution to the
$\rho\to4\pi$ decay amplitudes generated by the terms
Eq.~(\ref{rhorhopipi}). They are
\begin{eqnarray}
i\Delta
M^{(\rho\rho\pi\pi)}(\rho^0_q\to\pi^+_{q_1}\pi^+_{q_2}\pi^-_{q_3}\pi^-_{q_4})&=&
-i\frac{ag}{8f^2_\pi}(1+P_{12})(1+P_{34})(1+P_{24})(1-P_{13})\times\nonumber\\&&
\left[\frac{(1-P_{12})(\epsilon
q_1)(q_2q_4)}{D_\rho(q_1+q_3)}-\frac{(\epsilon
q_1)(q_2q_3)}{m^2_\rho}\right],\nonumber\\ i\Delta
M^{(\rho\rho\pi\pi)}(\rho^0_q\to\pi^+_{q_1}\pi^-_{q_2}\pi^0_{q_3}\pi^0_{q_4})&=&
-i\frac{ag}{8f^2_\pi}(1+P_{34})(1-P_{12})\left[(1-P_{13})\times\right.\nonumber\\&&\left.
\frac{(1-P_{14})(\epsilon
q_1)(q_2q_4)}{D_\rho(q_1+q_3)}-\frac{(\epsilon
q_1)(q_2q_3)}{m^2_\rho}\right],\nonumber\\ i\Delta
M^{(\rho\rho\pi\pi)}(\rho^-_q\to\pi^+_{q_1}\pi^-_{q_2}\pi^-_{q_3}\pi^0_{q_4})&=&
-i\frac{ag}{8f^2_\pi}(1+P_{23})\left[(1-P_{24})\frac{(1-P_{12})(\epsilon
q_1)(q_2q_3)}{D_\rho(q_2+q_4)}+\right.\nonumber\\&&\left.(1-P_{12})\frac{(1-P_{13})(\epsilon
q_1)(q_3q_4)}{D_\rho(q_1+q_2)}+(1-P_{14})\times\right.\nonumber\\&&\left.\frac{(1-P_{12})(\epsilon
q_1)(q_2q_3)}{D_\rho(q_1+q_4)}+(1-P_{24})(1+P_{13})
\right.\times\nonumber\\&&\left.\frac{(\epsilon
q_2)(q_1q_4)}{m^2_\rho}\right],\nonumber\\ i\Delta
M^{(\rho\rho\pi\pi)}(\rho^-_q\to\pi^-_{q_1}\pi^0_{q_2}\pi^0_{q_3}\pi^0_{q_4})&=&
i\frac{ag}{8f^2_\pi}(1+P_{23}+P_{24})(1-P_{12})(1+P_{34})\times\nonumber\\&&
\left[\frac{(1-P_{13})(\epsilon
q_1)(q_3q_4)}{D_\rho(q_1+q_2)}-\frac{(\epsilon
q_1)(q_2q_3)}{m^2_\rho}\right].\label{dem2rh2pi}\end{eqnarray} The
total amplitude for the decay $\rho\to4\pi$ is obtained upon
adding the pure HLS contribution $M^{(\rm HLS)}$ from
Refs.~\cite{ach00,ach03} and the above mentioned Eq.~(\ref{dela1})
(and similar expressions) together with
Eq.~(\ref{dem2rh2pi}):\begin{equation}M_{\rho\to4\pi}=M^{(\rm
HLS)}_{\rho\to4\pi}+\Delta M^{(a_1\rho\pi)}_{\rho\to4\pi}+\Delta
M^{(\rho\rho\pi\pi)}_{\rho\to4\pi}\equiv\epsilon_\mu
J_\mu(\rho\to\pi_{q_1}\pi_{q_2}\pi_{q_3}\pi_{q_4}).
\label{amp4pitot}\end{equation} The expressions for
$J_\mu(\rho\to\pi_{q_1}\pi_{q_2}\pi_{q_3}\pi_{q_4})$ are
excessively lengthy, even with the use of the permutation
operators $P_{ij}$, so we do not give them here.

The detailed analysis of the $\omega\to5\pi$ and  $\phi\to5\pi$
decay amplitudes is given elsewhere \cite{ach03}. As was shown
there, the $\rho\to4\pi$ transition amplitude  enters into the
dominant diagrams in Fig.~\ref{diagr_5pi}(a) corresponding to the
process $\omega,\phi\to\rho\pi\to5\pi$, in the following way:
\begin{equation}
M_{\omega_q,\phi_q\to5\pi}=\frac{g_{\rho\pi\pi}g_{\omega,\phi\to\rho\pi}}{f^2_\pi}
\varepsilon_{\mu\nu\lambda\sigma}q_\mu\epsilon_\nu\left[\frac{q_{5\lambda}
J_\sigma(\rho\to\pi_{q_1}\pi_{q_2}\pi_{q_3}\pi_{q_4})}{D_\rho(q-q_5)}+
\cdots\right],\label{ampv5pi}\end{equation} where particles are
labelled by their four-momentum, $\epsilon_\nu$ is the
polarization four-vector of the decaying $\omega,\phi$, and
$\cdots$ means the terms obtained from the written one by the
permutation of the pion momenta plus the contributions from the
remaining diagrams in Fig.~\ref{diagr_5pi} (b)$-$(e).
%%%%%%%%%%%%%%%%%%%%%%%%%%%%%%%%%%%%%%%%%%%%%%%%%%%%
\begin{figure}
\includegraphics{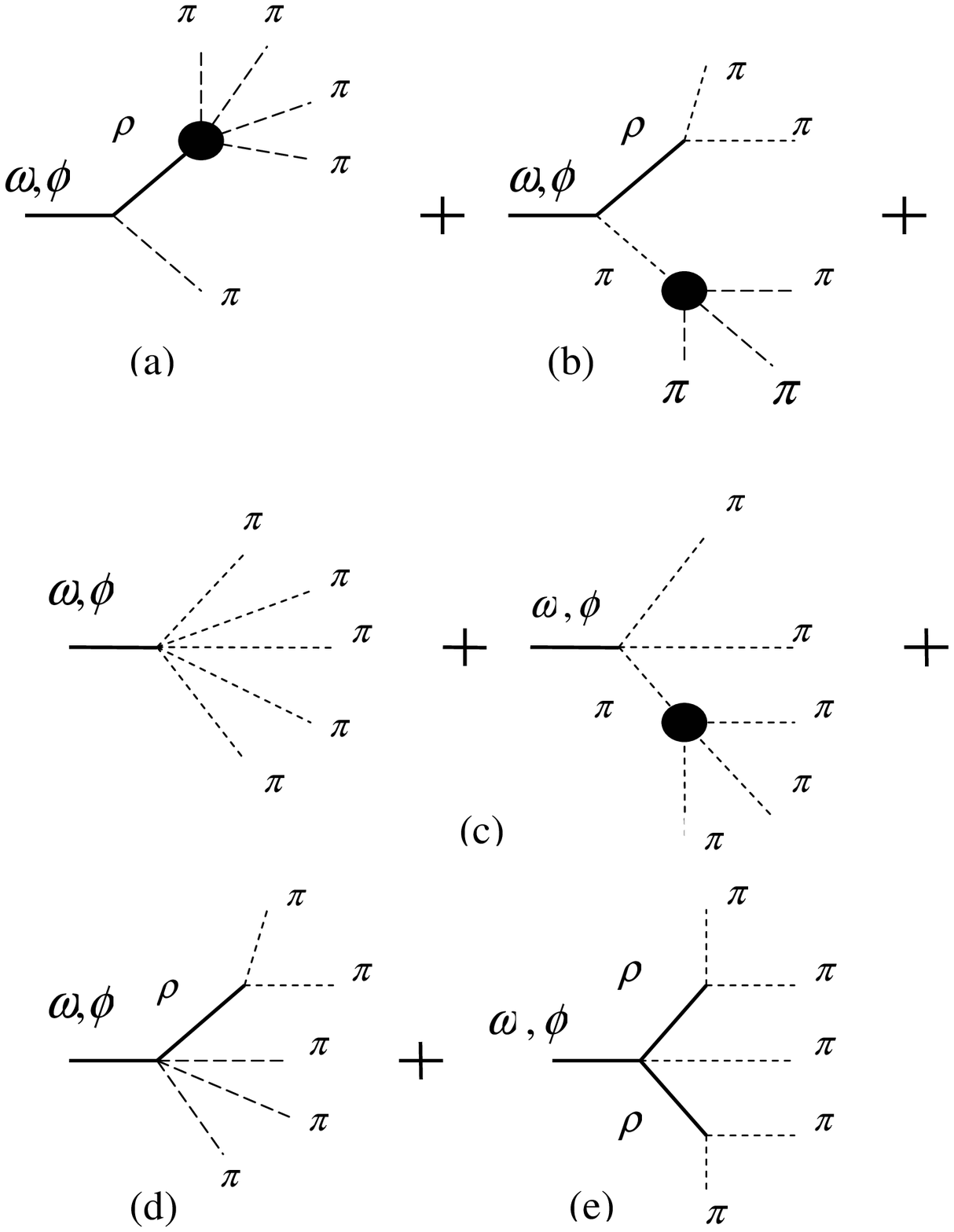}
\caption{\label{diagr_5pi}The schematic diagrams for  the
amplitude of the decay $\phi,\omega\to5\pi$.  The shaded circle in
(a), (b), (c) refers,respectively, to the $\rho\to4\pi$ transition
amplitude Eq.~(\ref{amp4pitot}) and $\pi\to3\pi$ vertex given in
Refs.~\cite{ach00,ach03}. }
\end{figure}
%%%%%%%%%%%%%%%%%%%%%%%%%%%%%%%%%%%%%%%%%%%%%%%%%
Taking into account the $a_1$ resonance in the generalized hidden
local symmetry approach reduces to the use of the total
$\rho\to4\pi$ decay current in that part of the
$\omega,\phi\to5\pi$ decay amplitude which corresponds to the
process $\omega,\phi\to5\pi$ with the resonant intermediate $\rho$
meson, see the diagram Fig.~\ref{diagr_5pi}(a). The latter term
means hereafter that the upper kinematical bound on the invariant
mass of the four-pion system in the final state of the decay
$\omega,\phi\to5\pi$ can be greater than the $\rho$ mass. As can
be seen  from Eq.~(\ref{ampv5pi}), the amplitude obeys the Adler
condition. Indeed, the contribution without the $a_1$ resonance
was shown  to obey this condition \cite{ach03}, while the
$a_1\rho\pi$ and $\rho\rho\pi\pi$ terms discussed earlier in this
paper satisfy this property separately. We do not give here the
explicit expressions for the full amplitudes because they are very
cumbersome.

\section{Branching ratios of the decays $\rho\to4\pi$, $\omega\to5\pi$, and
$\phi\to5\pi$ evaluated with the $a_1$ contribution
}\label{results} ~

Using Eq.~(\ref{amp4pitot}), the $\rho\to4\pi$ decay width is
evaluated according to the expression
\begin{eqnarray}
\Gamma_{\rho\to4\pi}(s)&=&\frac{1}{3\times\pi^6\times
s^{3/2}\times2^{12}\times
N_s}\int_{s_{1-}}^{s_{1+}}ds_1\int_{s_{2-}}^{s_{2+}}ds_2
\int_{u_{1-}}^{u_{1+}}\frac{du_1}{\lambda^{1/2}(s,s_2,s^\prime_2)}
\int_{u_{2-}}^{u_{2+}}du_2\times\nonumber\\&&\int_{-1}^1\frac{d\zeta_2}{(1-\zeta^2_2)^{1/2}}
|M_{\rho\to4\pi}(s,s_1,s_2,u_1,u_2,t_2(\zeta_2))|^2,
\label{widthrho4pi}\end{eqnarray} where the modulus squared of the
matrix element summed over the polarization states of the initial
$\rho$ meson,
$|M_{\rho\to4\pi}(s,s_1,s_2,u_1,u_2,t_2(\zeta_2))|^2$, is
expressed through the Mandelstam-like invariant variables $s=q^2$,
$s_ 1=(q-q_1)^2$, $s_2=(q_3+q_4)^2$, $u_1=(q-q_2)^2$,
$u_2=(q-q_3)^2$, $t_2=(q_1+q_4)^2$, $s^\prime_2=(q_1+q_2)^2$. See
Ref.~\cite{kumar}, where the expressions for the limits of
integration and $t_2\equiv t_2(\zeta_2)$ are given. The Bose
symmetry factor is $N_s=2$  for the decay modes
$\rho^0\to\pi^+\pi^-\pi^0\pi^0$,
$\rho^\pm\to\pi^\pm\pi^\pm\pi^\mp\pi^0$, $N_s=4$ for the mode
$\rho^0\to\pi^+\pi^-\pi^+\pi^-$, and $N_s=6$ for the decay mode
$\rho^\pm\to\pi^\pm\pi^0\pi^0\pi^0$. Notice that  the isotopic
mass difference of the charged and neutral pion is taken into
account both in the phase space volume and the decay matrix
element. The results of calculation are given in
Table~\ref{width4pi}, where the $\rho\to4\pi$ decay widths  are
presented in the cases of $m_{a_1}=m_\rho\sqrt{2}=1.09$ GeV (the
Weinberg relation), $m_{a_1}=1.23$ GeV (the PDG value \cite{pdg}),
and in case when the $a_1$ and counter term contributions are
neglected. One can see that the lower mass of the $a_1$ meson
results in a greater decay rate. This enhancement is due to the
low energy tail  of the $a_1$ Breit-Wigner factor.
%%%%%%%%%%%%%%%%%%%%%%%%%%%%%%%%%%%%%%%%%%%%
\begin{table}%[H] add [H] placement to break table across pages
\caption{\label{width4pi}The width of the decay $\rho\to4\pi$
[keV] evaluated in the model of generalized hidden local symmetry
\cite{bando88a}, at different masses of the $a_1$ resonance. The
uncertainty of the quoted central values  set to  about $10\%$ is
due to the difference in the value of gauge coupling constant
$g=g_{\rho\pi\pi}$ found from the $\rho\to\pi^+\pi^-$ decay width
or from KSRF relation Eq.~(\ref{ksrf}). }
\begin{ruledtabular}
\begin{tabular}{ccccc}
$m_{a_1}$
[GeV]&$\Gamma_{\rho^0\to2\pi^+2\pi^-}(m^2_\rho)$&$\Gamma_{\rho^0\to\pi^+\pi^-2\pi^0}(m^2_\rho)$&
$\Gamma_{\rho^\pm\to2\pi^\pm\pi^\mp\pi^0}(m^2_\rho)$&$\Gamma_{\rho^\pm\to\pi^\pm3\pi^0}(m^2_\rho)$\\
\hline 1.09&1.84&0.81&1.53&1.17\\1.23&1.59&0.75&1.38&1.00\\no
$a_1$&0.94&0.59&0.99&0.59
\end{tabular}
\end{ruledtabular}
\end{table}

The partial width of the decay
$\omega,\phi\to\pi_{q_1}\pi_{q_2}\pi_{q_3}\pi_{q_4}\pi_{q_5}$,
where the pions are labelled by their four-momenta,  is evaluated
according to the expression
\begin{eqnarray}
\Gamma_{\omega,\phi\to5\pi}(s)&=&\frac{\pi^2\sqrt{s}}{8\times3\times
N_{\rm
sym}\times(2\pi)^{11}}\int_{s_{1-}}^{s_{1+}}ds_1\int_{s_{2-}}^{s_{2+}}ds_2
\int_{s_{3-}}^{s_{3+}}ds_3\int_{u_{1-}}^{u_{1+}}du_1\times\nonumber\\&&\int_{u_{2-}}^{u_{2+}}
\frac{du_2}{[\lambda(s,s_2,s^\prime_2)\lambda(s,m^2_3,u_2)]^{1/2}}
\int_{u_{3-}}^{u_{3+}}
\frac{du_3}{[\lambda(s,s_3,s^\prime_3)\lambda(s,m^2_4,u_3)]^{1/2}}\times\nonumber
\\&&\int_{t_{2-}}^{t_{2+}}
\frac{dt_2}{[\lambda(s,t_1,t^\prime_1)(1-\xi^2_2)(1-\eta^2_2)(1-\zeta^2_2)]^{1/2}}
\times\nonumber
\\&&\int_{t_{3-}}^{t_{3+}}
\frac{dt_3|M_{\omega,\phi\to5\pi}(s,s_1,s_2,s_3,u_1,u_2,u_3,t_2,t_3)|^2}
{[\lambda(s,t_2,t^\prime_2)(1-\xi^2_3)(1-\eta^2_3)
(1-\zeta^2_3)]^{1/2}}, \label{gam5pi}\end{eqnarray}where
$s=(\sum_{a=1}^5 q_a)^2$; the Bose symmetry  factor is $N_{\rm
sym}=4$, 6 in case of the final state $2\pi^+2\pi^-\pi^0$,
$\pi^+\pi^-3\pi^0$, respectively. The basic integration variables
due to Kumar \cite{kumar} are
\begin{eqnarray}
s_1&=&(q-q_1)^2,\nonumber\\
s_2&=&(q-q_1-q_2)^2,\nonumber\\s_3&=&(q-q_1-q_2-q_3)^2,\nonumber\\
u_1&=&(q-q_2)^2,\nonumber\\u_2&=&(q-q_3)^2,\nonumber\\u_3&=&(q-q_4)^2,\nonumber\\
t_2&=&(q-q_2-q_3)^2,\nonumber\\t_3&=&(q-q_2-q_3-q_4)^2,
\label{kumarvar}\end{eqnarray}$t_1\equiv u_1$, $t^\prime_1\equiv
m^2_2$. The variables $s^\prime_2=(q_1+q_2)^2$,
$s^\prime_3=(q_1+q_2+q_3)^2$, $t^\prime_2=(q_2+q_3)^2$,
$\xi_{2,3}$, $\eta_{2,3}$, and $\zeta_{2,3}$ can be expressed
through the ones Eq.~(\ref{kumarvar}), see Ref.~\cite{kumar}. The
limits of integration in Eq.~(\ref{gam5pi}) are also given there.
$|M_{\omega,\phi\to5\pi}(s,s_1,s_2,s_3,u_1,u_2,u_3,t_2,t_3)|^2$ is
the modulus squared of the $\omega,\phi\to5\pi$ decay amplitude
summed over polarization states of the decaying particle. It
should be expressed through the same variables. The necessary
expressions of the scalar products $(q_aq_b)$, $a,b=1,\cdots5$ can
be found in Ref.~\cite{ach03}. The latter reference is devoted to
the evaluation of  the branching ratios of the decays
$\omega,\phi\to\pi^+\pi^-3\pi^0$ and
$\omega,\phi\to2\pi^+2\pi^-\pi^0$  in the  HLS scheme using the
lagrangian Eq.~(\ref{lan}) in the case of the $\omega(782)$ and
analogous lagrangian in the case of $\phi(1020)$. Taking the $a_1$
resonance into account in GHLS model reduces to using the total
$\rho\to4\pi$ decay current obtained in the previous section,  in
that part of the $\omega,\phi\to5\pi$ decay amplitude which
corresponds to the process $\omega,\phi\to\rho\pi\to5\pi$ with the
resonant intermediate $\rho$ meson.  See Eq.~(\ref{ampv5pi}).

As was pointed out in Ref.~\cite{ ach03}, the lagrangian
Eq.~(\ref{lan}) induced by the anomalous term of Wess and Zumino
(and analogous expression in the case of $\phi$) can be used for
the evaluation  of the $\omega,\phi\to5\pi$ decay rates only under
the definite  assumptions about arbitrary parameters $c_{1,2,3}$ (
and analogous parameters in the case of $\phi$). The choice
\begin{equation}
c_1=c_3\mbox{, }c_2=0\mbox{, }a=2.\label{sumrel}\end{equation}
made in Ref.~\cite{ach03} is used here, too. With this choice, the
$\omega\to5\pi$ decay rate is determined by the  coupling constant
Eq.~(\ref{gomrp}) only. The variation of $c_{1,2,3}$ within rather
wide margins around the values given by Eq.~(\ref{sumrel}) imply
no significant changes in the branching ratio. As for  the
$\phi\to5\pi$ decay, its branching ratio is determined within the
accuracy $20\%$ by the effective coupling constant
$g_{\phi\rho\pi}$ extracted from the $\phi\to\pi^+\pi^-\pi^0$
decay width. The results are insensitive to the choice of free
parameters analogous to Eq.~({\ref{sumrel}). See Ref.~\cite{ach03}
for the detailed study of this question. The results of the
evaluation are presented in Table \ref{width5pi}. Notice the
difference in the central value of
$B_{\phi\to2\pi^+2\pi^-\pi^0}=5.0\times10^{-7}$ in the lower line
of this table with the figure $(6.9\pm1.4)\times10^{-7}$ given in
Ref.~\cite{ach03}. This is due to the typesetting error in the
program code for  the non-leading contribution represented by the
anomaly induced terms corresponding to the process
$\rho\to\omega\pi\to4\pi$. Such terms refer to higher derivatives
in the effective lagrangian. The tail of this error disappears
upon the energy decrease. Indeed, the value of
$B_{\omega\to2\pi^+2\pi^-\pi^0}$ here and in Ref.~\cite{ach03}
differs by the factor 1.06.  The error is fixed when preparing the
present paper.
%%%%%%%%%%%%%%%%%%%%%%%%%%%%%
\begin{table}%[H] add [H] placement to break table across pages
\caption{\label{width5pi}The branching ratios  of the decays
$\omega(782)\to5\pi$ and $\phi(1020)\to5\pi$ evaluated in the
model of generalized hidden local symmetry \cite{bando88a} added
with the anomaly induced terms \cite{bando88}, at different masses
of the $a_1$ resonance. The uncertainty of the central values due
to the parameter dependence of the anomaly induced terms  is set
to $\pm20\%$, see Ref.~\cite{ach03}.}
\begin{ruledtabular}
\begin{tabular}{ccccc}
$m_{a_1}$
[GeV]&$B_{\omega\to\pi^+\pi^-3\pi^0}(m^2_\omega)$&$B_{\omega\to2\pi^+2\pi^-\pi^0}(m^2_\omega)$&
$B_{\phi\to\pi^+\pi^-3\pi^0}(m^2_\phi)$&$B_{\phi\to2\pi^+2\pi^-\pi^0}(m^2_\phi)$\\
\hline
1.09&$4.2\times10^{-9}$&$3.8\times10^{-9}$&$4.4\times10^{-7}$&$8.8\times10^{-7}$\\
1.23&$4.1\times10^{-9}$&$3.7\times10^{-9}$&$3.9\times10^{-7}$&$7.7\times10^{-7}$\\no
$a_1$&$3.6\times10^{-9}$&$3.3\times10^{-9}$&$2.5\times10^{-7}$&$5.0\times10^{-7}$
\end{tabular}
\end{ruledtabular}
\end{table}
%%%%%%%%%%%%%%%%%%%%%%%%%%%%%%%%%%%%%%%%%%%%%%%%%%%%%%%%%%%%%%%%%%%%%%%%%%%%%%%%%%%%%%%%
%%%%%%%%%%%%%%%%%%%%%%%%%%%%%%%%%%%%%%%%%%%%%%%%%%%%
\begin{figure}
\includegraphics{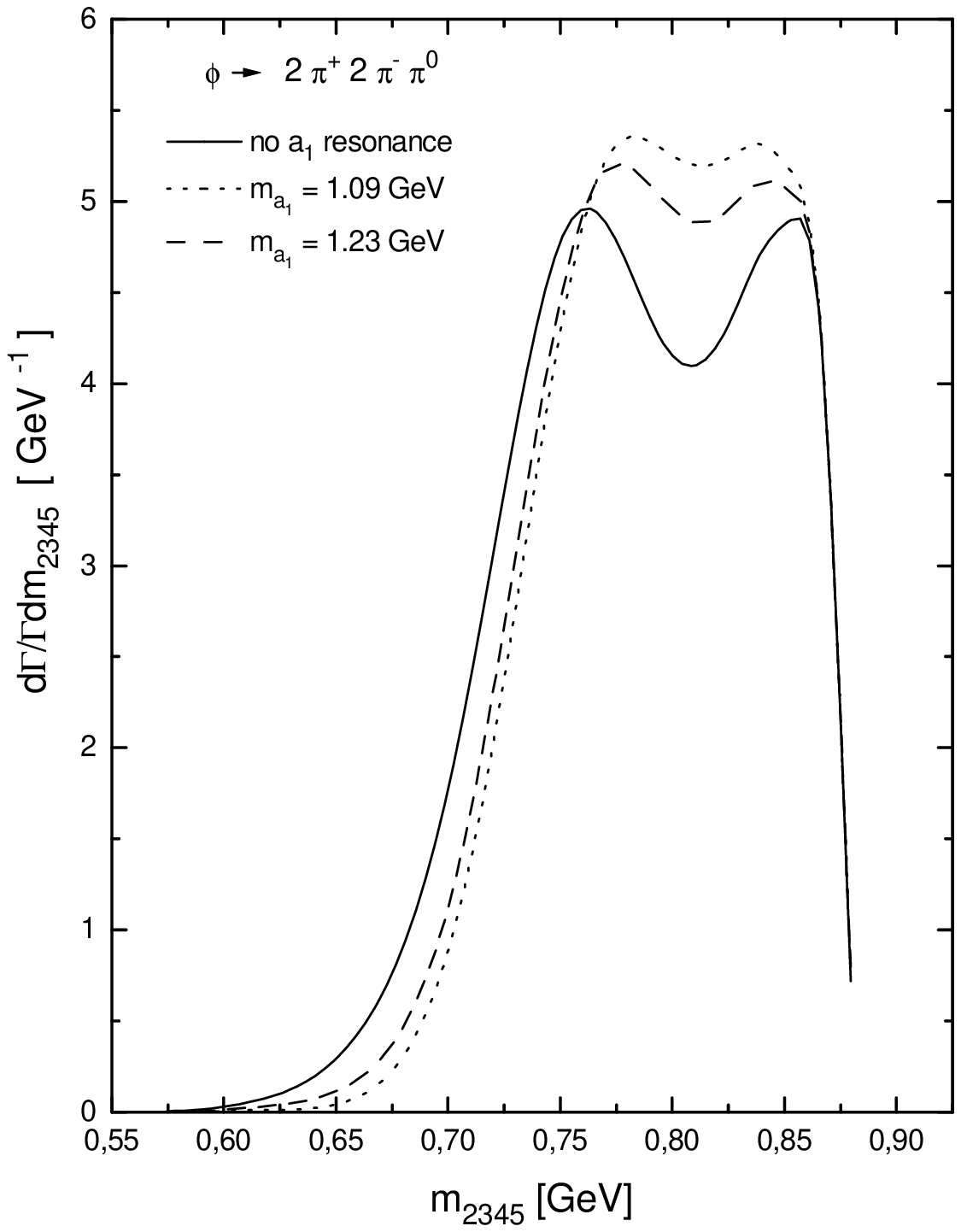}
\caption{\label{spec5pic}The mass spectrum of the system
$\pi^+_{q_2}\pi^-_{q_3}\pi^-_{q_4}\pi^0_{q_5}$ in the decay
$\phi\to\pi^+_{q_1}\pi^+_{q_2}\pi^-_{q_3}\pi^-_{q_4}\pi^0_{q_5}$
normalized to the respective $5\pi$ decay width, and calculated at
$\sqrt{s}=m_\phi$. The invariant mass squared is
$m^2_{2345}=(q_2+q_3+q_4+q_5)^2\equiv s_1$.}
\end{figure}
%%%%%%%%%%%%%%%%%%%%%%%%%%%%%%%%%%%%%%%%%%%%%%%%%
%%%%%%%%%%%%%%%%%%%%%%%%%%%%%%%%%%%%%%%%%%%%%%%%%%%%
\begin{figure}
\includegraphics{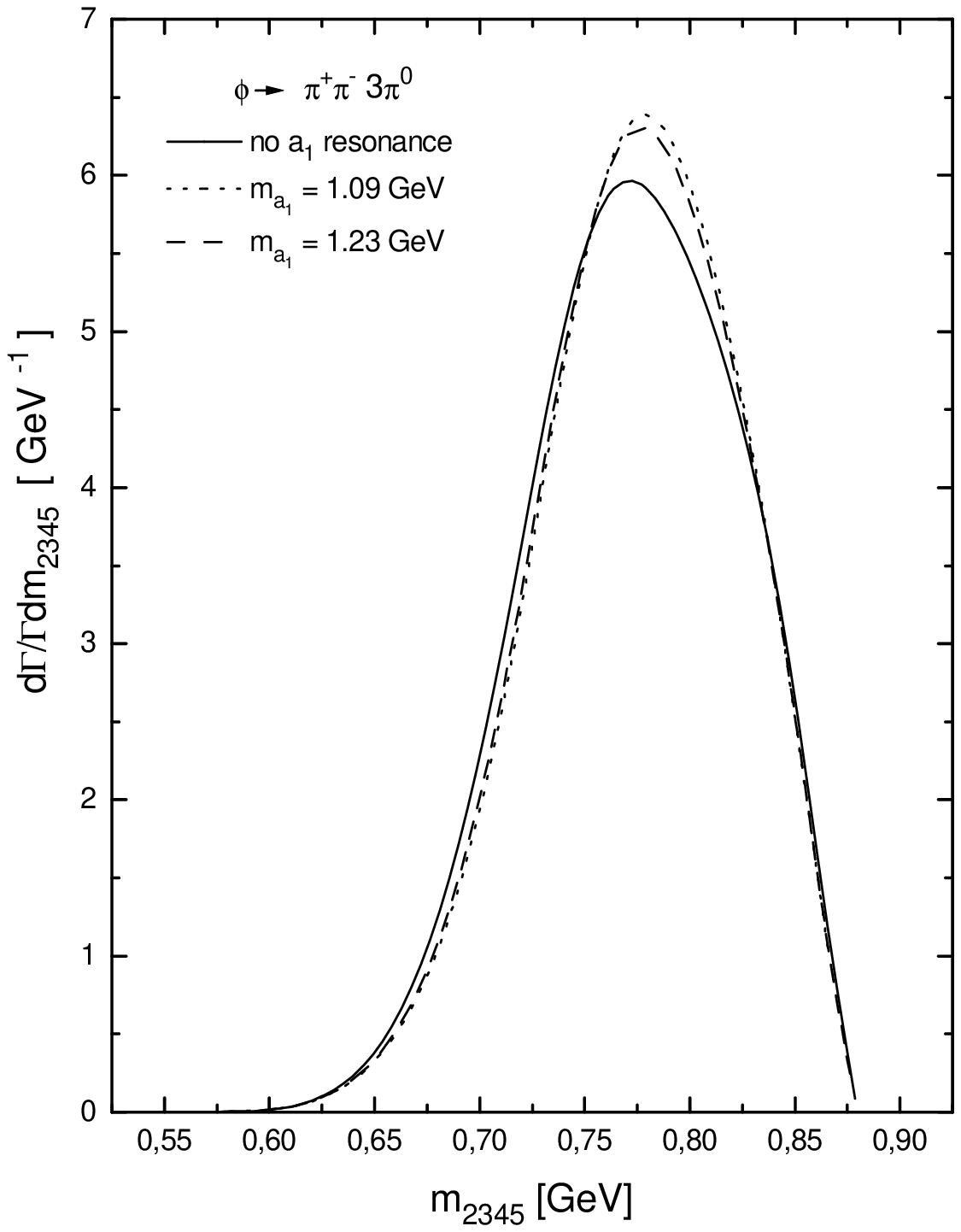}
\caption{\label{spec5pin}The mass spectrum of the system
$\pi^-_{q_2}\pi^0_{q_3}\pi^0_{q_4}\pi^0_{q_5}$ in the decay
$\phi\to\pi^+_{q_1}\pi^-_{q_2}\pi^0_{q_3}\pi^0_{q_4}\pi^0_{q_5}$
normalized to the respective $5\pi$ decay width, and calculated at
$\sqrt{s}=m_\phi$. The invariant mass squared is
$m^2_{2345}=(q_2+q_3+q_4+q_5)^2\equiv s_1$.}
\end{figure}
%%%%%%%%%%%%%%%%%%%%%%%%%%%%%%%%%%%%%%%%%%%%%%%%%

It is interesting to plot the mass spectrum of the four pion
subsystem in the final five-pion state. This can be fulfilled
straightforwardly  for the distribution over the Kumar variable
$\sqrt{s_1}$, see Eq.~(\ref{kumarvar}). The corresponding curves
for the decays $\phi\to2\pi^+2\pi^-\pi^0$ and
$\phi\to\pi^+\pi^-3\pi^0$ are shown in Fig.~\ref{spec5pic} and
\ref{spec5pin} in the cases when, first, no $a_1$ resonance is
present and, second,  the $a_1$ resonance is included with the
above chosen masses $m_{a_1}=1.09$ and 1.23 GeV. The spectra look
different. Specifically, both spectra has the peak due to the
$\rho$ pole. In the meantime, the mass spectrum of the subsystem
$\pi^+\pi^-\pi^-\pi^0$ possesses the second  peak, while the
$\pi^-\pi^0\pi^0\pi^0$ one does not. This is due to the presence
of the strong energy dependent anomaly induced contribution
$\rho^-\to\omega\pi^-\to\pi^+\pi^-\pi^-\pi^0$ in the decay
$\phi\to\rho^-\pi^+\to\pi^+\pi^+\pi^-\pi^-\pi^0$ followed by the
necessary phase space kinematical cut-off. There is no anomaly
induced enhancement in the decay
$\phi\to\rho^-\pi^+\to\pi^+\pi^-\pi^0\pi^0\pi^0$. The
distributions over  invariant mass of the remaining four-pion
subsystems,  $\sqrt{u_{1,2,3}}$ and $\sqrt{s^\prime_5}$, where
$s^\prime_5=(q-q_5)^2$ \cite{kumar}, in principle, can be obtained
upon inserting the corresponding $\delta$ function into
Eq.~(\ref{gam5pi}). In practice, however, this demands  the
complex rearrangements of the sequential integration bounds  in
Eq.~(\ref{gam5pi}) \cite{kumar}. We do not make this task here.
Instead, we  restrict ourselves by drawing the qualitative
conclusions that the mass spectra of the subsystems
$\pi^+\pi^-\pi^0\pi^0$ and $\pi^+\pi^-\pi^+\pi^-$ should look
similar to ones shown in Fig.~\ref{spec5pic} and \ref{spec5pin},
respectively, because in the former, like in the plotted
$\pi^+\pi^-\pi^-\pi^0$ one, there is also the anomaly induced
contribution $\rho^0\to\omega\pi^0\to\pi^+\pi^-\pi^0\pi^0$ while
in the latter one there is no such contribution, see
Ref.~\cite{ach00} for more detail. To summarize the discussion of
the mass spectrum of the four pion subsystem, we point out that,
as is evident from Fig.~\ref{spec5pic} and \ref{spec5pin}, the
greater part of the total number of events of the decay
$\phi\to5\pi$ should originate from the
$\pm\frac{1}{2}\Gamma_\rho$ vicinity of the $\rho$ peak in the
process $\phi\to\rho\pi\to5\pi$.

\section{Discussion}
\label{discussion}

Let us compare the part of  our results concerning the widths of
the decays $\rho^0\to2\pi^+2\pi^-$ and $\rho^0\to\pi^+\pi^-2\pi^0$
with those of Ref.~\cite{plant96}. One can see that our
calculation in cases when the $a_1$ resonance is taken into
account, gives the partial widths which exceed those obtained in
Ref.~\cite{plant96} by a factor ranging from 1.5 to 1.8, depending
on the mass of the $a_1$ meson. In the meantime, our calculation
gives the coinciding results in the model without $a_1$ meson.
When making such a comparison, note, first, that here we take into
account the mass difference of the charged and neutral pions both
in matrix elements and the phase space volume, while the authors
of Ref.~\cite{plant96} set all pion masses equal to the mass of
$\pi^\pm$. Second, we fix $g_{\rho\pi\pi}$ from the
$\rho\to\pi^+\pi^-$ width while in Ref.~\cite{plant96} it is fixed
by Eq.~(\ref{ksrf}). The mentioned difference between the results,
in all appearance, could be attributed to the way of taking into
account the contribution of the $a_1$ resonance. Indeed, as is
discussed in Sec.~\ref{a1width}, there are different ways of
taking into account the additional terms arising due to the
diagonalization of the axial-pseudoscalar mixing. This could
result in the terms similar to Eq.~(\ref{a13piadd}), which, in
principle, could affect the specific value of the $\rho\to4\pi$
width. Unfortunately, the authors of Ref.~\cite{plant96} did not
give the necessary details to make the comparison and reveal the
reason of the discussed discrepancy.

The KLOE collaboration at DA$\Phi$NE $\phi$ factory has collected
the total number of events at $\sqrt{s}=m_\phi$ equivalent to the
luminosity integral $\int{\cal L}dt\approx500$ pb$^{-1}$
\cite{kloe}. Using the table \ref{width5pi}, one can estimate the
expected number of events $N_{\phi\to5\pi}$ of the decay
$\phi\to5\pi$ which already could be present in the whole KLOE
statistics. One obtains $N_{\phi\to5\pi}\approx 1340$, 2070, 2360,
respectively, in the HLS model without $a_1$ meson, in the GHLS
model which incorporates the $a_1$ meson with the    mass
$m_{a_1}=1.23$ GeV, 1.09 GeV.

\begin{acknowledgements}
The present study was partially supported by the grants
RFFI-02-02-16061 from Russian Foundation for Basic Research and
NSh-2339.2003.2 for Support of Leading Scientific Schools.
\end{acknowledgements}

%%%%%%%%%%%%%%%%%%%%%%%%%%%%%%%%%%%%%%%%%%%%%%%%%%%%%%%%%%


\begin{thebibliography}{99}
\bibitem{fn1}
Note that the second of the diagrams Fig.~\ref{diagr_5pi}(c) does
not contribute, provided the condition Eq.~(\ref{rel1}) is
satisfied. However, we keep here the total set of the diagrams,
because it makes possible to study the effect of the relaxing the
above constrain and, in turn, to estimate the accuracy of the
model dependence upon the free parameters $c_{1,2,3}$  in the case
of the $\omega\to5\pi$ decay and the analogous parameters in the
case of the $\phi\to5\pi$ decay. See Ref.~\cite{ach03} for more
detail.
\end{thebibliography}
\end{document}